Microcomb-referenced photonic stabilization of resonant tunneling diode terahertz oscillators

Miezel Talara[1,†], Yu Tokizane[1,2,†], Tatsunoshin Mori[3], Ryota Shikata[3], Masayuki Higaki[3], Eiji Hase[1,2], Isao Morohashi[4], Safumi Suzuki[5], Naoya Kuse[1,2], and Takeshi Yasui[1,2]

[1]Institute of Post-LED Photonics (pLED), Tokushima University, 2-1, Minami-Josanjima, Tokushima 770-8506, Japan

[2]Institute of Photonics and Human Health Frontier (IPHF), Tokushima University, 2-24, Shinkura, Tokushima, Tokushima, 770-8501, Japan

[3]Graduate School of Sciences and Technology for Innovation, Tokushima University, 2-1, Minami-Josanjima, Tokushima 770-8506, Japan

[4]National Institute of Information and Communications Technology, 4-2-1 Nukuikitamachi, Koganei, Tokyo 184-8795, Japan

[5]Institute of Integrated Research, Institute of Science Tokyo, 2-12-1 Ookayama, Meguro-ku, Tokyo 152-8552, Japan

[†]These authors contributed equally.

**ABSTRACT**

We demonstrate a compact stabilization scheme for terahertz (THz) sources by exploiting the complementary advantages of microresonator-based optical frequency combs (microcombs) and resonant tunneling diodes (RTDs). A microcomb-driven photomixing THz signal is employed as the master for injection locking of an RTD, enabling faithful transfer of the microcomb stability into the RTD. Using this approach, the free-running RTD linewidth of 50 MHz was narrowed to 165 Hz, while the single-sideband phase noise reached -80 dBc/Hz at a 10 kHz offset with a locking range of 80 MHz. Compared with conventional electronic frequency multiplier or fiber-comb-based schemes, this method avoids high-order frequency multiplication and associated noise penalties, offering a compact and practical alternative. The dual functionality of linewidth narrowing and power scalability highlights the potential of microcomb-assisted injection locking as a route toward chip-scale, spectrally pure THz sources for beyond-5G/6G wireless communication and radar, with prospects for future extension to time-frequency metrology and precision sensing.

# 1. INTRODUCTION

Resonant tunneling diode (RTD) oscillators are regarded as promising compact sources in the terahertz (THz) band [1,2]. They combine room-temperature operation, small device dimensions, and compatibility with semiconductor processes, making them attractive for integrated and low-cost THz systems. These features align well with the growing demand for communication [3-5], radar [6-8], and sensing [9]. Unlike frequency-multiplier-based sources, RTDs directly generate THz radiation from micrometer-scale devices, providing advantages in efficiency and integration. However, free-running RTDs suffer from broad linewidths on the order of several to hundreds of megahertz, reflecting high phase noise induced by device instabilities and environmental fluctuations. Such limited spectral purity might be problematic for several applications; for instance, it is incompatible with highly advanced modulation formats in high-capacity wireless links, where phase fluctuations degrade error vector magnitude and bit error rate, and it also limits range resolution in radar systems. Therefore, while RTDs possess unique strengths as compact THz oscillators, their inherent phase instability remains a barrier to wider practical use. Narrowing the linewidth and reducing phase noise are indispensable for establishing RTDs as viable sources for THz applications that require high spectral purity.

Phase-locked loop (PLL) techniques have been applied to stabilize RTD oscillators [10,11], but their use at THz frequencies is limited by narrow loop bandwidths, circuit complexity, and poor scalability, as well as by the reliance on such as electronic frequency-multiplier-based reference sources. Injection locking offers a simpler alternative, providing passive synchronization of frequency and phase without wideband feedback [12]. Two main approaches have been reported in THz region. In the first approach, RTDs were injection-locked to electronic frequency multipliers as the master source, achieving precise frequency control and linewidths of a few hundred kilohertz around 382 GHz [13]. However, this approach inevitably suffers from phase-noise degradation ($\Delta L_{FM}$) associated with frequency multiplication, which is given by

$$\Delta L_{FM} = 20 log_{10}(M) \tag{1}$$

where *M* is the multiplication factor. For example, when generating a 300-GHz signal from a microwave synthesizer via multiplication, an 11.1-GHz reference with a 27× multiplier experiences about +28.6 dB degradation. In addition, such systems require bulky and power-hungry synthesizers and amplifiers, hindering compact and scalable integration. In the second approach, a combination of a mode-locked fiber laser-based optical frequency comb (ML-FL-OFC) and photomixing was employed as the master source to transfer the excellent stability of the comb repetition rate ($f_{rep}$) into the THz domain, achieving relative linewidths at the sub-hertz level around 332 GHz when heterodyne detection was performed using a photonic local oscillator phase-locked to the same ML-FL-OFC [14]. Yet this approach also relies on high-order frequency multiplication in optical region. For instance, stabilizing the $f_{rep}$ of a ML-FL-based OFC with a 100 MHz RF reference corresponds to an effective 3000× multiplication, which raises the phase-noise floor by nearly +69.5 dB in the 300 GHz band. Although the use of sub-hertz-linewidth lasers locked to high-finesse optical cavities as frequency references can suppress this noise and yield ultra-low-phase-noise carriers [15], the associated systems remain bulky, costly, and complex, and the dense comb spectra require additional filtering to extract isolated line pairs. These drawbacks emphasize the need for practical master sources that avoid high-order multiplication, thereby achieving low phase noise.

Microresonator-based OFCs (microcombs) have recently attracted significant attention as new OFC sources [16-18]. By exploiting the Kerr nonlinearity in chip-scale microresonators, they can generate broadband, widely spaced optical lines while maintaining compatibility with CMOS fabrication. Compared with ML-FL-based OFCs, microcombs offer clear advantages in footprint, simplicity, and potential for large-scale photonic-electronic integration, making them highly attractive. A particularly important feature is that their repetition rate, $f_{rep}$, can reach hundreds of gigahertz, directly within the THz domain, owing to the orders-of-magnitude smaller cavity size compared with conventional OFC sources. This allows neighboring comb lines to be heterodyned through photomixing to generate THz waves without any optical frequency multiplication, thereby avoiding the severe phase-noise degradation associated with high-order frequency multiplication in ML-FL-OFC-based approaches. Recent demonstrations have reported low-phase-noise THz generation through photomixing of microcombs [19-21], along with their applicability in coherent wireless links operating [22-24]. Such results highlight the promise of microcomb-based photonic THz sources as practical enablers of narrow-linewidth, low-phase-noise THz waves. However, the intrinsically low output power of photonic THz sources remains a major limitation for many practical applications. These demonstrations motivate further exploration of how such photonic THz sources can be harnessed to actively stabilize electronic oscillators. Building on these advances, the present work further explores microcomb-driven photomixing as a master signal for injection locking of RTD oscillators, aiming to combine the stability of microcombs with the power efficiency of RTDs for next-generation THz applications.

In this work, we employ microcomb-driven photomixing THz wave as the master signal for injection locking of RTD oscillators. By using THz waves generated via photomixing of free-running microcombs or phase-noise-suppressed microcombs with a uni-traveling carrier photodiode (UTC-PD) [25,26], we verify that the $f_{rep}$ of the microcomb can be faithfully transferred into the THz domain without degradation. This approach directly addresses the limitations of conventional electronic and ML-FL-OFC-based schemes by avoiding high-order frequency multiplication while maintaining a compact and practical architecture. Furthermore, it leverages the complementary characteristics of RTDs and photomixing: RTDs provide relatively high output power but suffer from broad linewidths and large phase noise, whereas photomixing of microcombs yields highly stable but low-power THz waves. By combining these attributes through injection locking, the RTD oscillator can operate as a spectrally purified source. This dual functionality underscores the potential of microcomb-assisted injection locking as a practical route toward compact and spectrally pure THz sources, ultimately bridging the gap between device-level demonstrations and real-world applications.

## 2. EXPERIMENTAL SETUP

2.1 Soliton microcomb generation

In our experiments, soliton microcombs were generated using a custom SiN ring microresonator (LIGENTEC, S.A., free spectral range ≈ 283 GHz, Q ≈ $10^6$) using a direct fiber-to-microresonator coupling scheme [27], as illustrated in Fig. 1(a). The fiber-coupled microresonator employed high-numerical-aperture (high-NA) polarization-maintaining single-mode fibers (PMFs), single-core fiber arrays (FAs), and a UV-curable optical adhesive to ensure efficient and mechanically stable coupling of the pump laser beam to the microresonator. Soliton formation was facilitated by rapid wavelength sweeping of the pump laser [28]. The output light of a single-mode CW laser (Santec TSL-550, linewidth = 200 kHz) passed through a dual-parallel Mach-Zehnder modulator (DP-MZM, T.SBXH1.5-35PD-ADC, Sumitomo Osaka Cement Co., Ltd., data rate = 40 Gbps), and then was amplified by an erbium-doped fiber amplifier (EDFA1, EYDFA-C-HP-BA-30-PM-B, CivilLasers, wavelength = 1535-1565 nm, saturated output power = 1 W). The resulting laser light (optical power = 1 W) was coupled into the fiber-coupled microresonator as a pump light. To generate a single soliton microcomb, the DP-MZM rapidly tunes the optical frequency of the pump light across the resonant frequency of the microresonator. The generated soliton microcomb was subsequently passed through an optical band-stop filter (OBSF) to remove the residual pump light and then amplified to an optical power of over 30 mW by another erbium-doped fiber amplifier (EDFA2, EDFA100P, Thorlabs., wavelength = 1530-1556 nm, saturated output power = 100 mW).

Figure 1(b) illustrates the setup for reducing the phase noise of the $f_{rep}$, which employs the two-wavelength delayed self-heterodyne interferometer (TWDI) [21,29,30] as a reference. The TWDI approach generates two microcomb replicas with a relative frequency shift and time delay, and their interference provides a direct measure of phase fluctuations in $f_{rep}$. The TWDI technique is not only a powerful tool for evaluating phase noise but can also be used actively for phase-noise suppression. In this configuration, a portion of the generated soliton microcomb is directed into a fiber-based Mach-Zehnder interferometer. One arm contains an acousto-optic modulator (AOM) that shifts the optical frequency by 80 MHz, producing a frequency-shifted microcomb, while the other arm incorporates a spool fiber to introduce a time delay and thus generates a time-delayed microcomb. These two signals are combined using a fiber coupler (FC) to produce optical interference, and adjacent comb line pairs generate an 80 MHz optical beat signal. A pair of bandpass filters (OBPFs) selects beat signals from two specific microcomb line pairs, which are then detected by photodetectors (PDs). The resulting 80 MHz beat signal is electrically mixed down to DC, which directly reflects the $f_{rep}$ phase noise. By feeding this signal as an error signal through a servo filter to control the optical power of the pump laser, $f_{rep}$ is phase-locked to the TWDI reference, thereby achieving low-phase-noise operation. The sensitivity and response speed of this control loop are determined by the length of the spool fiber and the comb line indices ($m$) used for detection. In the present work, a spool fiber length of 100 m was used, and comb lines at $m$ = +2 ($\lambda$ = 1555.4 nm) and $m$ = -3 ($\lambda$ = 1544.2 nm) were employed.

2.2 RTD injection locking

Figure 2 shows the experimental setup for RTD injection locking.

**Master source:** The master source was the microcomb-driven photomixing transmitter. The microcomb output was first split by a fiber coupler (FC) into two arms, one for the TWDI and the other for THz generation via photomixing. In the photomixing arm, the optical power was adjusted to 30 mW using a variable optical attenuator (ATT), after which the light was launched into a uni-traveling carrier photodiode (UTC-PD; NTT Electronics, IOD-PMJ-13001, operating range = 280-330 GHz, output power = -7 dBm at 283 GHz) integrated with a horn antenna (HA1; Virginia Diodes, RCHO3R, 220-325 GHz, gain = 21.5 dBi at 283 GHz). Through photomixing of the beat components generated by multiple pairs of adjacent microcomb lines, a continuous-wave THz wave (freq. = 283 GHz, output power ≈ 100 µW) was radiated from the HA1 into free space with vertical linear polarization (polarization angle or PL angle = 0°). The radiated THz wave (indicated in orange) was collimated by a THz lens (L1; diameter = 30 mm, focal length = 50 mm), passed through a wire-grid polarizer (WG; Origin, MWG40FA-II, freq. = 0.1-1.4 THz, extinction ratio = 7000:1, transmission axis = 0°), and subsequently focused by another THz lens (L2; diameter = 30 mm, focal length = 30 mm) onto the RTD device via a second horn antenna (HA2; lab-made).

**Slave source and detection:** The slave source, RTD oscillator integrated with the HA2, emitted a THz wave (freq. ≈ 283 GHz, output power ≈ 13 µW) with linear polarization oriented at 45°. This radiation (indicated in blue) was collimated by L2 and directed toward the WG. Because the 90° polarization component of the RTD emission was reflected by the WG, it was focused by a THz lens (L3; diameter = 30 mm, focal length = 50 mm) and subsequently coupled into a waveguide-integrated sub-harmonic mixer (SHM; Virginia Diodes, WR3.4MixAMC-I, RF freq. = 220-330 GHz, LO freq. = 9.17-13.75 GHz, IF freq. = DC-40 GHz) equipped with a third horn antenna (HA3; Virginia Diodes, WR-3.4, 220-330 GHz, gain = 21 dBi at 283 GHz). In the SHM, the reflected THz wave was down-converted by mixing with a local oscillator (LO) signal derived from a microwave synthesizer (Anritsu MG36021A, freq. = 11.7 GHz, output power = 0 dBm) followed by an 8× frequency multiplier (FM1; Virginia Diodes, WR9.0 AMC-I, freq. = 82-125 GHz) and a subsequent 3× multiplier (FM2; Virginia Diodes, WR2.8X3UHP, freq. = 250-375 GHz), yielding a total multiplication factor of 24. The resulting intermediate-frequency (IF) signal (freq. = 2 GHz) was characterized using two instruments: the power spectrum was measured with a spectrum analyzer (Anritsu, MS2840A-044), while the phase-noise spectrum was measured with a signal analyzer (Agilent, N9000B).

**Injection locking procedure:** For injection locking, the bias current of the RTD oscillator was carefully tuned to adjust its free-running oscillation frequency. By sweeping the current, the RTD emission frequency was aligned such that the photomixing-derived master THz wave fell within the locking range of the RTD. Under this condition, stable injection locking of the RTD oscillator to the microcomb-driven photonic THz source was successfully achieved.

## 3. RESULTS

3.1 Injection locking range of an RTD oscillator using a free-running microcomb

We first evaluated the frequency pulling and locking range of the RTD oscillator using a free-running microcomb-driven photomixing THz wave as the master signal. The THz wave generated from the UTC-PD was fixed at 283 GHz with an output power of 100 µW, while the oscillation frequency of the RTD was tuned by varying the applied bias voltage from 336 mV to 345 mV. Figure 3 shows the evolution of the RTD emission spectrum under different conditions, without injection, outside the locking range, and inside the locking range (RBW = 300 kHz, sweep time = 14 ms). The black trace shows both the free-running RTD emission and the photomixed THz wave that is reflected from the RTD toward the detector via WG. Because the injected photomixed THz wave lies outside the locking range, no injection locking occurs, and the two signals appear independently in the spectrum. As the RTD oscillation frequency is swept toward the injected frequency, the blue and cyan traces exhibit distorted spectra: the original RTD emission collapses into a mixture of a spike-like feature corresponding to the injected THz frequency and a broad background on the lower-frequency side. This behavior indicates partial pulling without complete locking. The green, yellow-green, and yellow traces represent the fully locked state, in which only the spike-like spectral component remains and the broad background disappears. This clearly demonstrates passive phase locking of the RTD to the injected signal (injection-locked RTD, IL-RTD). The orange trace shows a near-locking condition, where weak residual background is still visible but the spike-like component dominates. When the RTD frequency is further detuned beyond the locking range, as seen in the red and olive traces, the spectrum again consists of spike-like features mixed with a broad background. These behaviors confirm the characteristic spectral evolution associated with injection locking. We defined the locking range as the frequency interval over which the RTD emission peak remained at least 10 dB above the noise floor. Based on this definition and spectral measurements, the locking range was determined to be 80 MHz.

3.2 THz spectral comparison of free-running and injection-locked RTDs

We next compared the spectral characteristics of the RTD with and without injection locking (IL). Without IL, the free-running RTD exhibited a linewidth of approximately 50 MHz, as shown in Fig. 4(a) (RBW = 30 kHz, sweep time = 263 ms). This broad linewidth reflects the intrinsic characteristics of the free-running RTD. The linewidth of RTD oscillators generally is known to narrow with increasing output power owing to enhanced oscillation stability; thus, given that the RTD in this experiment operated at a relatively low output power, the observed linewidth is reasonable. In contrast, IL-RTD spectrum was significantly narrowed as shown in Fig. 4(b) (RBW = 300 Hz, sweep time = 14 ms). The linewidth was reduced to below 410 Hz, corresponding to a reduction by a factor of $1.2 \times 10^5$ compared with the non-IL case, thus clearly demonstrating the linewidth-narrowing effect of IL. However, even under IL, when the RTD was injection-locked by the free-running microcomb, the RF spectrum showed noticeable frequency drift (not shown).

We then evaluated linewidth narrowing by IL using a photomixed THz wave generated from a TWDI-stabilized microcomb as the master source. Figure 4(c) shows the spectrum of the IL-RTD (RBW = 100 Hz, sweep time = 41 ms) when injection-locked to the photomixed THz wave from the stabilized microcomb. Compared with the free-running-microcomb case in Fig. 4(b), where the linewidth was 410 Hz, the linewidth was further narrowed to 165 Hz. This demonstrates short-term linewidth narrowing by IL by a factor of approximately 2.5. Moreover, the frequency drift observed with the free-running microcomb was effectively suppressed under TWDI stabilization (not shown). The quantitative suppression of frequency drift is investigated in the following subsection.

3.3 Frequency drift and Allan deviation analysis of injection-locked RTDs

To quantitatively evaluate how effectively the TWDI stabilization of the microcomb suppresses THz-frequency drift, we monitored the temporal frequency fluctuations of injection-locked RTDs using free-running and TWDI-stabilized microcombs as master sources. Figure 5(a) shows the temporal evolution of the THz frequency deviation, where the frequency at the start of the measurement is defined as zero. The blue trace corresponds to the IL-RTD driven by a free-running microcomb. In this case, a pronounced frequency drift on the order of a few megahertz is observed over time. The standard deviation of the frequency fluctuation in this condition is 388 kHz, indicating that residual phase noise and drift of $f_{rep}$ in the free-running microcomb are directly transferred to the THz domain. In contrast, the red trace in Fig. 5(a) shows the frequency deviation of the IL-RTD driven by a TWDI-stabilized microcomb. The large frequency drift observed in the free-running case is strongly suppressed. Figure 5(b) presents an expanded view of the stabilized case, revealing residual frequency fluctuations on the order of several tens of kilohertz. The standard deviation of the frequency fluctuation is reduced to 17.4 kHz, corresponding to a suppression of frequency instability to approximately 4.5% of the free-running case.

To further quantify the frequency stability over different time scales, we evaluated the Allan deviation of the THz frequency, as shown in Fig. 5(c). The Allan deviation was calculated for gate times ranging from 1 ms to 10 s. The blue and red plots correspond to IL-RTDs driven by free-running and TWDI-stabilized microcombs, respectively. Across this range of gate times, the TWDI stabilization yields a frequency fluctuation suppression of more than one to two orders of magnitude compared with the free-running case. At gate times longer than approximately 100 ms, the Allan deviation for the stabilized case exhibits an increasing trend. This behavior suggests that the long-term stability is currently limited by the time stability of the TWDI reference and by non-optimized feedback control parameters, rather than by an intrinsic limitation of the injection-locking approach. In particular, further improvement of the TWDI time stability will require enhanced thermal control and increased robustness against environmental disturbances such as vibration or sounde. With such improvements, together with optimization of the TWDI loop bandwidth and control conditions, further enhancement of long-term frequency stability is expected.

3.4 Phase-noise comparison between the master THz wave and the injection-locked RTD using a TWDI-stabilized microcomb

Finally, in order to confirm that the low-phase-noise characteristics of the stabilized $f_{rep}$ signal were faithfully transferred to the IL-RTD without degradation, we evaluated the phase noise performance. Figure 6 shows the single-sideband (SSB) phase noise power spectral density (PSD) as a function of offset frequency, i.e., the phase noise spectrum. The blue trace represents the photomixed THz wave generated from the stabilized microcomb (master source), while the red trace corresponds to the IL-RTD. The two phase-noise spectra overlap closely over an offset-frequency range from 100 Hz to 100 kHz, indicating that the phase-noise characteristics of the master THz wave are effectively transferred to the RTD through injection locking within this bandwidth. At offset frequencies above 100 kHz, however, a deviation between the master and IL-RTD spectra becomes apparent. Such behavior has been reported in previous studies, where incomplete phase-noise tracking occurs when the injected master power is insufficient relative to the free-running oscillator power, resulting in partial rather than ideal phase-noise following of the master signal [31]. The observed discrepancy at higher offset frequencies in the present measurements is therefore attributed to this finite injection ratio between the master photomixed THz wave and the RTD oscillator.

For comparison, the estimated phase-noise spectrum of the LO signal used in the SHM is shown by the green trace in Fig. 6. This spectrum was estimated from the phase-noise specification of the microwave synthesizer output at 10 GHz, to which the effect of 24-fold frequency multiplication ($24^2$ = 576, corresponding to a +27.6 dB increase in phase noise) was added to obtain the equivalent THz phase-noise level. In addition, the phase-noise floor of the signal analyzer is indicated by the black trace, which is based on the instrument specification at 1 GHz, chosen as the closest available reference to the IF frequency band (~2 GHz). From this comparison, three important conclusions can be drawn. First, the observed phase-noise characteristics are not limited by the phase noise of the LO used for the SHM or by the noise floor of the measurement instruments. Second, the phase-noise characteristics of the microcomb-driven photomixing master THz wave are faithfully transferred to the RTD without degradation through injection locking. Third, a phase noise of -80 dBc/Hz at a 10 kHz offset frequency is achieved for the IL-RTD using the TWDI-stabilized microcomb. We note that further improvement of the phase-noise performance toward the -100 dBc/Hz level, comparable to prior state-of-the-art demonstrations [21], is expected to be feasible through further optimization of the TWDI-referenced $f_{rep}$ stabilization scheme.

## 4. DISCUSSION

This work demonstrates a new approach for stabilizing RTD oscillators by injection locking to a microcomb-driven photomixed THz wave, achieving both narrow linewidth and low phase noise within a practical architecture. Compared with ML-FL-OFCs, the advantage of microcombs can be quantified in terms of phase-noise scaling. Here, let us consider the case in which an ML-FL-OFC with a $f_{rep}$ of 100 MHz is phase-locked to a 10-MHz rubidium atomic clock (for example, -155 dBc/Hz phase noise at a 10-kHz offset). In this configuration, the $f_{rep}$ is stabilized by phase-locking it to a 100-MHz frequency synthesizer referenced to the atomic clock. According to Eq. (1), the frequency multiplication from 10 MHz to 100 MHz introduces a +20 dB degradation, and the subsequent multiplication from 100 MHz to 283 GHz (×2830) adds another +69 dB, resulting in a total penalty of +89 dB. As a result, the 283 GHz carrier inherits phase noise on the order of -66 dBc/Hz at 10 kHz offset. Alternatively, high-performance ML-FL-OFCs referenced to optical frequency standards can achieve much lower phase noise, because the process involves both optical frequency division and subsequent frequency multiplication [15]. The phase-noise reduction by frequency division ($\Delta L_{FD}$) is expressed as

$$\Delta L_{FD} = -20log_{10}(N) \qquad (2)$$

where $N$ is the division factor. For example, a cavity-stabilized ultranarrow-linewidth laser with -94.7 dBc/Hz at 10 kHz offset at 193.4 THz transfers its stability to $f_{rep}$ via a frequency division by 1.93×10$^8$ ($\Delta L_{FD}$ = -125 dB), followed by a multiplication to 283 GHz ($\Delta L_{FM}$ = +69 dB), leading to an estimated -150.7 dBc/Hz at 10 kHz offset.

Although this represents a substantial phase-noise advantage, it requires bulky, costly, and complex cavity-stabilized laser. By contrast, the present work achieves -80 dBc/Hz at a 10 kHz offset using a microcomb-driven photomixing source without reliance on large-scale optical references, by exploiting adjacent comb-line pairs and thereby avoiding phase-noise degradation induced by frequency multiplication. This quantitative comparison highlights the unique impact of microcomb-based injection locking in simultaneously providing miniaturization and spectral purity, pointing toward practical deployment in compact THz transceivers, frequency references, and chip-scale integrated systems.

While this study employed an RTD with a relatively modest output power of 13 µW, state-of-the-art devices have already demonstrated milliwatt-class emission from a single RTD oscillator [2]. In this context, the present approach can also be regarded as a hybrid scheme that combines the photonic stability of microcombs with the electronic power scalability of RTDs, effectively enabling the RTD to act as an optically driven THz amplifier [31]. Ongoing advances in RTD technology, including device scaling and array integration [2], suggest the potential for even higher output powers in the future. It should be noted, however, the pursuit of higher output power and broader locking range are inherently in trade-off. The theoretical locking range ($\Delta f_{lock}$) is given by [13]:

$$\Delta f_{lock} \propto \frac{f_0}{Q}\sqrt{\frac{P_{inj}}{P_{out}}} \tag{3}$$

where $f_0$ is the oscillation frequency of the RTD, $Q$ is the quality factor of the RTD oscillator, $P_{inj}$ is the injected master power of UTC-PD-photomixed THz wave received by the RTD, and $P_{out}$ is the output power of the RTD. In the present experiment, an RTD with an output power of 13 µW achieved a locking range of 80 MHz. For higher-power RTDs, maintaining a sufficiently wide locking range will require careful design to balance stability and power enhancement. Nevertheless, by leveraging high-power RTDs within the injection-locking framework demonstrated here, it becomes feasible to realize milliwatt-level THz waves that retain the low phase-noise properties of the microcomb master, offering a promising path toward practical THz applications that demand both high output power and high spectral purity.

To further clarify the positioning of our approach, it is instructive to compare it with other state-of-the-art stabilization schemes for RTD oscillators. Recent work based on injection locking of waveguide-integrated RTDs using dual-wavelength Brillouin lasers has demonstrated extremely low phase-noise characteristics, highlighting their potential as high-coherence THz sources [31]. However, these systems rely heavily on bulky and non-integrable waveguide components (WR-band mixers, isolators, and circulators) as well as Brillouin laser systems, and therefore face intrinsic limitations in scalability, particularly toward frequencies beyond 1 THz, where waveguide hardware becomes prohibitively difficult to realize. In contrast, the microcomb/UTC-PD/RTD architecture employed in this work is fully compatible with free-space or on-chip photonic coupling, avoids waveguide-dependent constraints, and offers superior scalability in frequency, footprint, and architectural flexibility. This distinction has important implications for long-term technological development: all key building blocks of our scheme, microcombs, UTC-PDs, and RTDs, already exist in chip-integrated implementations. Thus, future progress can leverage hybrid photonic-electronic integration to drastically reduce footprint, power consumption, and system complexity while enhancing robustness. Such chip-scale architectures are highly attractive for realizing compact and low-noise THz transmitters, and continued advances in RTD device technology offer a realistic prospect of achieving higher output power within this integrated framework.

## 5. CONCLUSION

This work has demonstrated a new stabilization strategy for THz sources by exploiting the complementary advantages of microcombs and RTDs. By injection locking RTDs to microcomb-driven photomixing THz signals, we verified that the low phase noise of $f_{rep}$ in the microcomb can be faithfully transferred to the RTD, achieving a linewidth narrowing from 50 MHz to 165 Hz and a phase noise of -80 dBc/Hz at a 10 kHz offset, with a locking range of 80 MHz. The approach distinguishes itself from ML-FL-OFC systems by offering a practical solution that simultaneously addresses the demands for miniaturization, low phase noise, and scalable output power. Importantly, even without absolute frequency referencing, such low-phase-noise THz carriers are highly valuable for applications including coherent wireless communication and radar.

While such stabilization for phase noise reduction is already valuable for communication and radar, certain applications, such as THz clocks for 6G test and measurement platforms, require absolute frequency referencing. In this study, the microcomb repetition rate was stabilized using the TWDI method, effectively suppressing relative phase noise. However, the absolute frequency was not defined. Establishing an absolute THz reference will require linking the microcomb to external standards. Possible directions include hybrid configurations that connect the microcomb to low-noise RF or microwave references via electro-optic-modulator-based combs [20] or a Vernier cavity [32], although achieving both phase locking to a reference and simultaneous phase-noise suppression by TWDI remains technically demanding. Such developments would enable the realization of absolute THz clocks, expanding the impact of microcomb-assisted RTD oscillators into time-frequency metrology and high-precision sensing.

Looking ahead, the proposed architecture offers a foundation for a new class of hybrid photonic-electronic THz sources. Because all key building blocks, microcombs, UTC-PDs, and RTDs, already exist in chip-integrated forms, future work can move beyond device-level demonstrations toward monolithically integrated THz systems. Such integration will enable compact, power-efficient, and spectrally pure THz modules that can be incorporated directly into practical instruments and networks. In this sense, the present demonstration not only validates microcomb-assisted stabilization of RTDs but also opens a pathway toward a unified platform in which photonic stability and electronic power scalability are inherently combined.

**ACKNOWLEDGEMENTS**

This work was supported in part by the Ministry of Internal Affairs and Communications (MIC) of Japan through the FORWARD program (JPMI240910001) and the R&D project "High-speed THz communication based on radio and optical direct conversion" (JPJ000254). Additional support was provided by Japan Society for the Promotion of Science (JSPS) program for Forming Japan's Peak Research Universities (J-PEAKS, Grant No. JPJS00420240022), the Cabinet Office of the Government of Japan (Promotion of Regional Industries and Universities), Tokushima Prefecture (Creation and Application of Next-Generation Photonics), and the Research Clusters Program of Tokushima University (2201001). We would like to express our sincere gratitude to ROHM Co., Ltd. for providing the resonant tunneling diode (RTD) used in this research. We also gratefully acknowledge Anritsu Corp. for providing the low-phase-noise microwave synthesizer.

## AUTHOR DECLARATIONS

### Conflict of Interest

The authors have no conflicts to disclosure interests.

### Author Contributions

**Miezel Talara**: Data curation (equal); Methodology (equal); Writing - review & editing (equal). **Yu Tokizane:** Conceptualization (equal); Data curation (equal); Methodology (equal); Investigation (equal); Writing - review & editing (equal); Supervision (equal). **Tatsunoshin Mori:** Data curation (equal); Methodology (equal); Investigation (supporting); Writing - review & editing (supporting).

**Ryota Shikata:** Data curation (supporting); Methodology (supporting); Investigation (supporting).

**Masayuki Higaki:** Data curation (supporting); Methodology (supporting). **Eiji Hase:** Supervision (supporting). **Isao Morohashi:** Methodology (supporting). **Safumi Suzuki:** Conceptualization (supporting); Methodology (supporting). **Naoya Kuse:** Conceptualization (equal); Methodology (supporting); Investigation (equal); Supervision (equal). **Takeshi Yasui:** Conceptualization (lead); Methodology (equal); Investigation (equal); Writing - original draft (lead); Writing - review & editing (equal); Supervision (lead); Funding acquisition (lead).

**DATA AVAILABILITY**

The data that support the findings of this study are available from the corresponding author upon reasonable request.

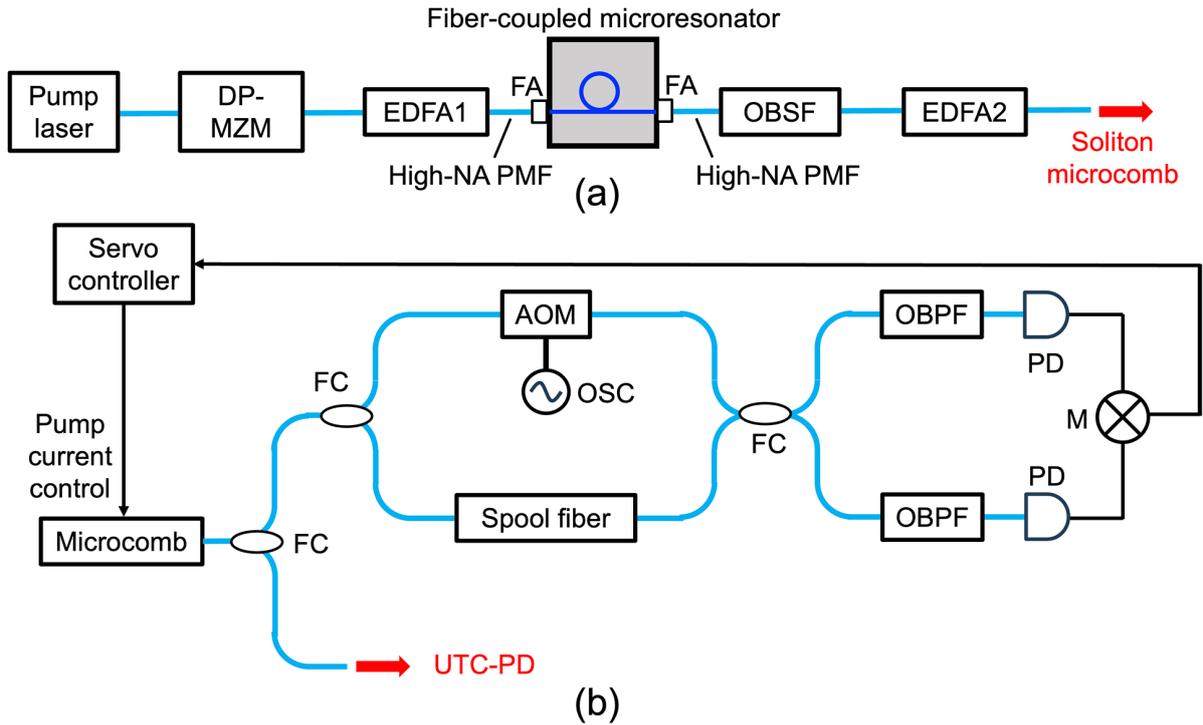

Fig. 1. Experimental setup for (a) soliton microcomb generation and (b) reducing the phase noise of $f_{rep}$. DP-MZM, dual-parallel Mach-Zehnder modulator; EDFA1 and EDFA2, erbium-doped fiber amplifiers; high-NA PMFs, high-numerical-aperture polarization-maintaining single-mode fibers; FAs, single-core fiber arrays ; OBSF, optical band-stop filter; FCs, fiber couplers; AOM, acousto-optic modulator; OSC, oscillator; OBPFs, optical band-pass filters; PDs, photodetectors; M, electric mixer; UTC-PD, uni-traveling carrier photodiode.

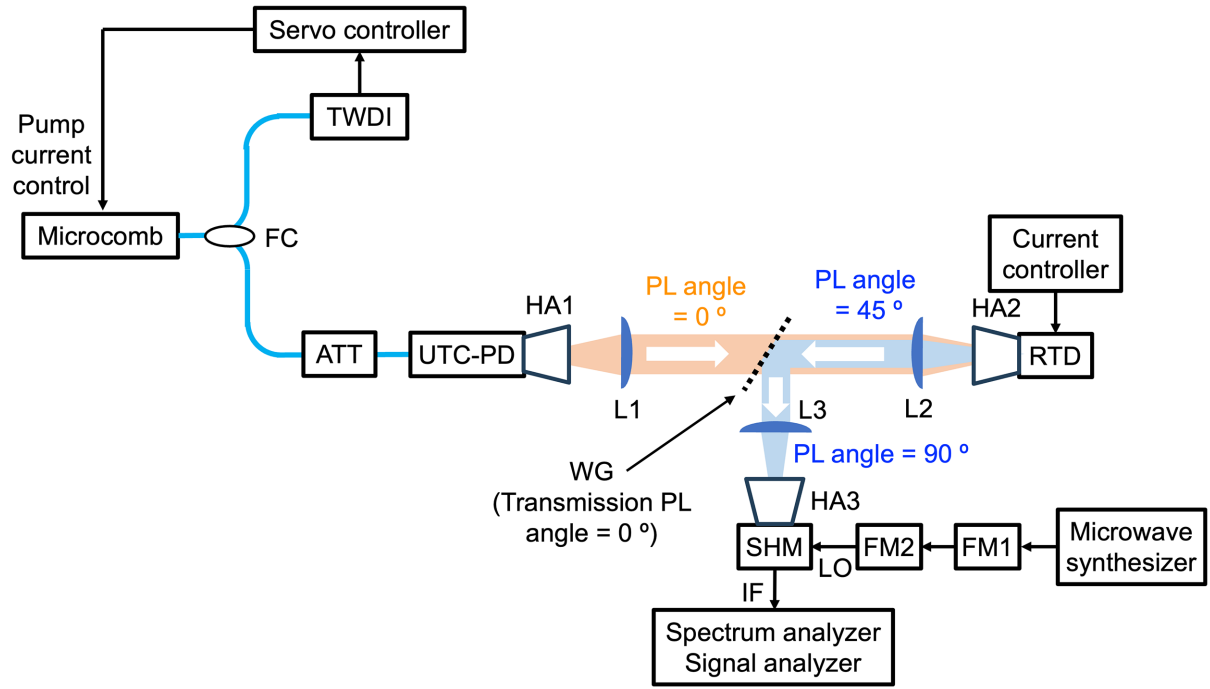

Fig. 2. Experimental setup for RTD injection locking. FCs, fiber couplers; TWDI, two-wavelength delayed self-heterodyne interferometer; ATT, variable optical attenuator; UTC-PD, uni-traveling carrier photodiode; HA1, HA2, HA3, horn antennas; L1, L2, L3, THz lenses; WG, wire-grid polarizer; RTD, resonant tunnel diode; SHM, waveguide-integrated sub-harmonic mixer; FM1 and FM2, frequency multipliers.

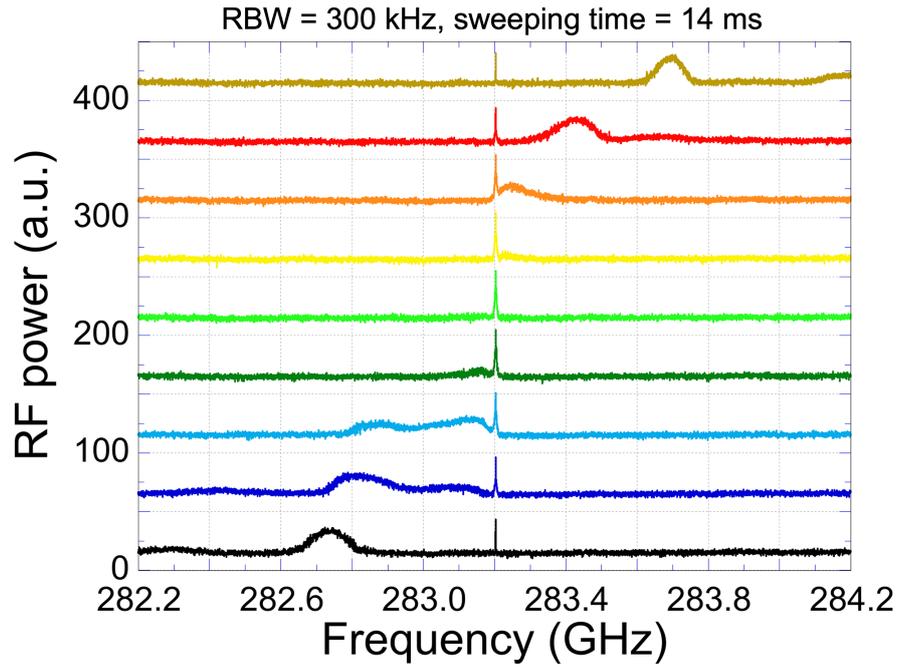

Fig. 3. Evolution of the RTD emission spectrum under different injection conditions (RBW = 300 kHz, sweep time = 14 ms). The black trace corresponds to the free-running RTD with the reflected photomixed THz wave outside the locking range. The blue and cyan traces show partial frequency pulling without complete locking. The green, yellow-green, and yellow traces indicate the fully injection-locked state (IL-RTD). The orange trace represents a near-locking condition. The red and olive traces correspond to detuned conditions outside the locking range.

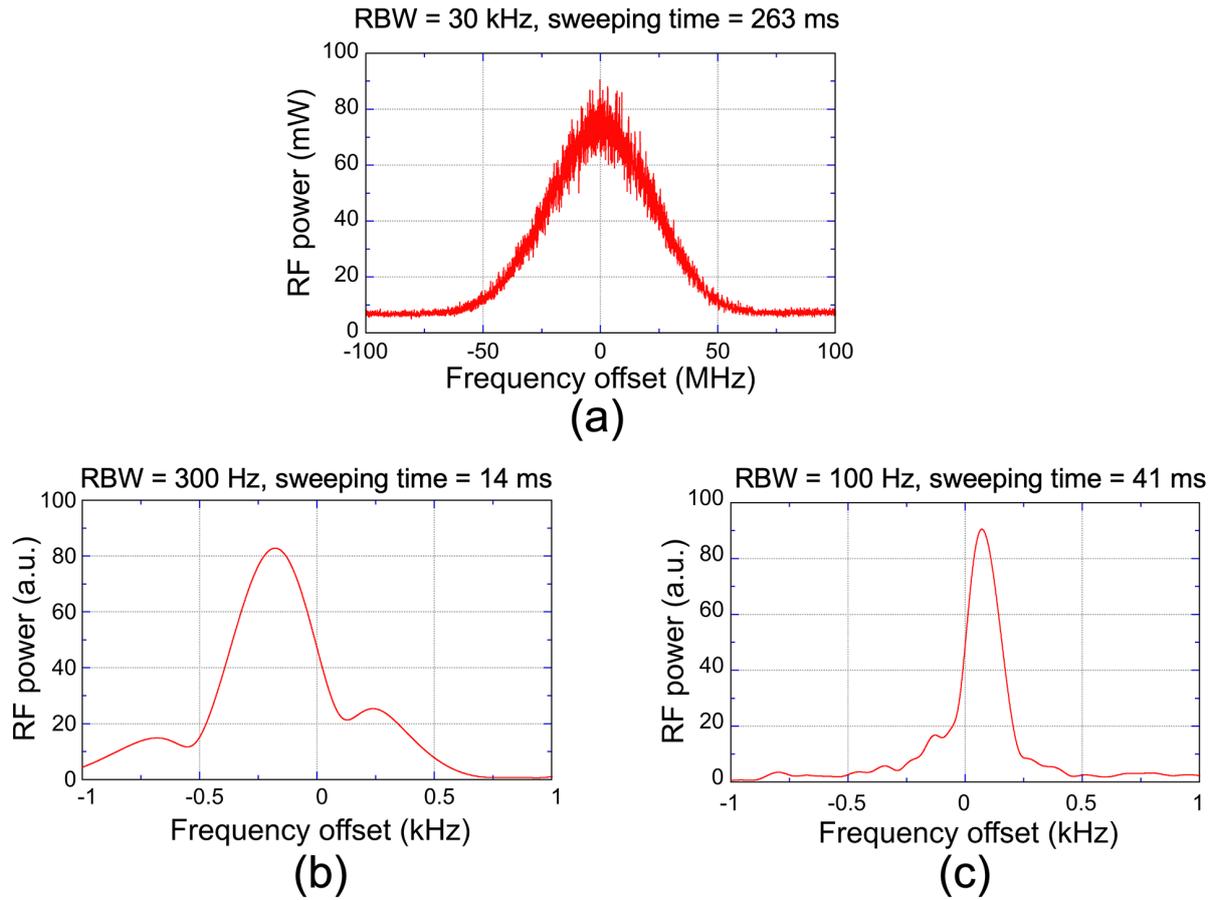

Fig. 4. Spectral characteristics of the RTD with and without injection locking. (a) Free-running RTD without injection locking, (b) injection-locked RTD (IL-RTD) using a free-running microcomb-driven photomixing THz wave, and (c) IL-RTD using a TWDI-stabilized microcomb-driven photomixing THz wave.

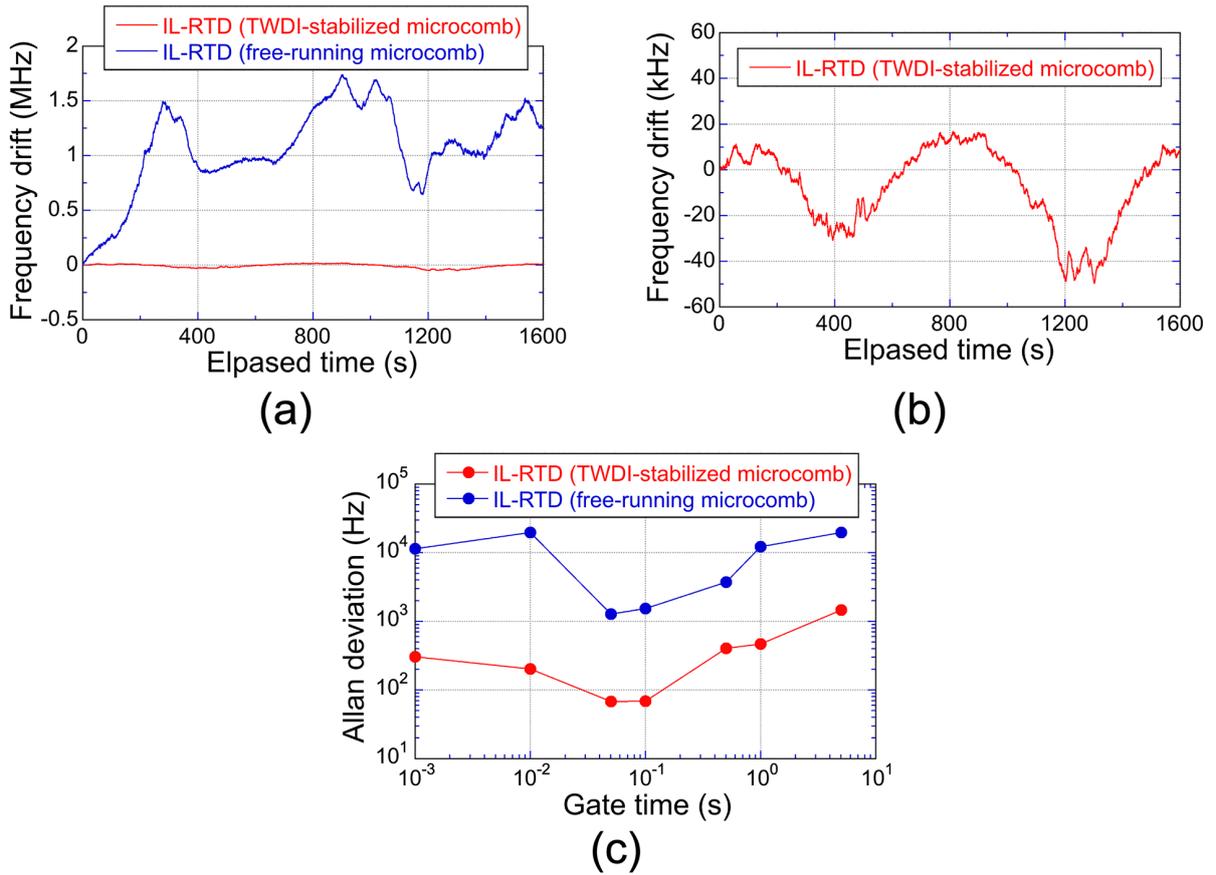

Fig. 5. Frequency stability comparison of injection-locked RTDs (IL-RTDs) using free-running and TWDI-stabilized microcomb-driven photomixing THz waves. (a) Frequency drift of the IL-RTD referenced to a free-running microcomb (blue) and a TWDI-stabilized microcomb (red) as a function of elapsed time. (b) Expanded view of the frequency drift for IL-RTD using the TWDI-stabilized microcomb. (c) Allan deviation of the IL-RTD for a free-running microcomb (blue) and a TWDI-stabilized microcomb (red).

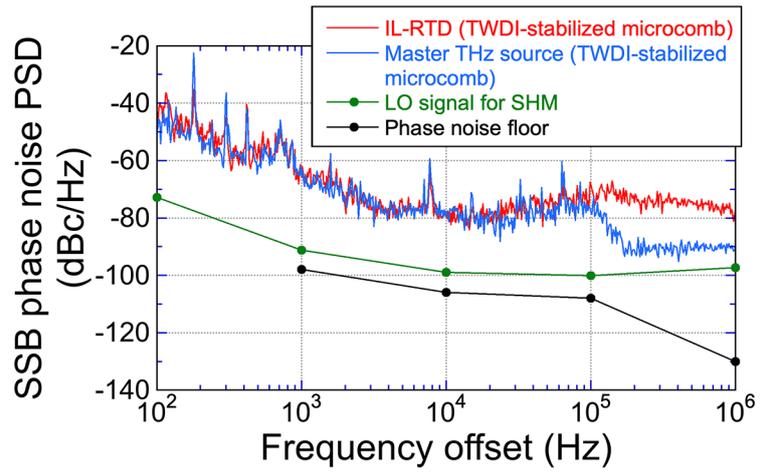

Fig. 6. Comparison of phase noise spectra among injection-locked RTD (TWDI-stabilized microcomb), master THz source (TWDI-stabilized microcomb), LO signal for SHM, and phase noise floor.